\documentclass[a4paper,aps,prd,onecolumn,preprintnumbers,showpacs,superscriptaddress,nofootinbib]{revtex4}

\usepackage{graphics,graphicx,epsfig}
\usepackage{amsfonts,amsmath,amssymb}

\begin{document}

\title{Shadow of a Kaluza--Klein rotating dilaton black hole}
\author{Leonardo Amarilla}
\email{yellow@df.uba.ar} \affiliation{Departamento de F\'{\i}sica, Facultad de Ciencias Exactas y Naturales, Universidad de Buenos Aires, Ciudad Universitaria, Pabell\'on 1, 1428, Buenos Aires, Argentina.}
\author{Ernesto F. Eiroa} \email{eiroa@iafe.uba.ar} \affiliation{Instituto de Astronom\'{\i}a y F\'{\i}sica del Espacio, Casilla de Correo 67, Sucursal 28, 1428, Buenos Aires, Argentina.} \affiliation{Departamento de F\'{\i}sica, Facultad de Ciencias Exactas y Naturales, Universidad de Buenos Aires, Ciudad Universitaria, Pabell\'on 1, 1428, Buenos Aires, Argentina.}


\begin{abstract}
We study the shadow produced by a spinning Kaluza--Klein black hole in Einstein gravity coupled to a Maxwell field and a dilaton. The size and the shape of the shadow depend on the mass, the charge, and the angular momentum. We find that, for fixed values of these parameters, the shadow is slightly larger and less deformed than for its Kerr--Newman counterpart.
\end{abstract}

\maketitle

\section{Introduction}

The study of the strong deflection gravitational lensing of light by compact objects has received great attention in recent years, motivated by considerable evidence of the presence of supermassive black holes at the galactic centers. The relevant astrophysical quantities, i.e. the positions, magnifications, and time delays of the relativistic images produced by black hole lenses, can be approximately obtained in an analytic way by using the strong deflection limit  for spherically symmetric lenses \cite{darwin,otros,eiroto,boz}. The method relies on a logarithmic expansion of the deflection angle for light rays passing close to the photon sphere. Black hole lensing was also numerically studied \cite{numerical}. Non-rotating black holes coming from alternative theories were considered as gravitational lenses as well \cite{alternative,bwlens}. Kerr black holes were analyzed as lenses by several researchers \cite{bozza1,bozza2,vazquez,kraniotis}; in particular, the strong deflection limit was extended to these
objects \cite{bozza1,bozza2}. The shadows (or apparent shapes) of non-rotating black holes are circular, but rotating ones present a deformation caused by the spin \cite{bardeen,chandra}. This topic has been recently considered by several authors  \cite{falcke,devries,takahashi,bozza2, hioki,bambi,maeda,amarilla1,zakharov,schee1,schee2,amarilla2,yumoto}, both in Einstein gravity and in modified theories, since it is expected that direct observation of black holes will be possible in the near future \cite{zakharov,morris,lacroix}. In that case, the analysis of the shadows will be a useful tool for obtaining properties of astrophysical black holes and comparing different gravitational theories. For a review of strong deflection lensing and its consequences see, for example, Ref. \cite{bozzareview}.

The action corresponding to standard gravity coupled to the Maxwell field $F^{\mu\nu}$ and the (scalar) dilaton field $\phi$, in geometrized units ($16\pi G=c=1$), reads \cite{mae,ghs,horne}
\begin{equation}
 S=\int d^4x\sqrt{-g}\left[ -R+2(\nabla \phi)^2+e^{-2\gamma \phi }F^2\right] ,
\label{emd}
\end{equation}
where $R$ is the Ricci scalar and $g$ is the determinant of the metric of the spacetime. This action can be obtained, in the Einstein frame, within the context of low energy string theory with a Maxwell field, but with  all other gauge fields and the antisymmetric field set to zero. The parameter $\gamma $ is included in the coupling between the dilaton and the Maxwell field, as is done in Refs. \cite{ghs,horne}; when $\gamma =0$ the Einstein--Maxwell scalar lagrangian is recovered. The action (\ref{emd}) leads to the Einstein equations with the dilaton and the Maxwell fields as the sources:

\begin{equation}
\nabla _{\mu }\left( e^{-2\gamma \phi }F^{\mu \nu }\right) =0,
\end{equation}
\begin{equation}
\nabla ^{2}\phi +\frac{\gamma }{2}e^{-2\gamma \phi }F^{2}=0,
\end{equation}
\begin{equation}
R_{\mu \nu }=2\nabla _{\mu }\phi \nabla _{\nu }\phi +2e^{-2\gamma \phi }
\left( F_{\mu \zeta }F_{\nu }^{\; \zeta }-\frac{1}{4}g_{\mu \nu }F^{2} \right) .
\label{feq}
\end{equation}
The static spherically symmetric solutions of the field equations were obtained in Refs. \cite{mae,ghs} for arbitrary values of $\gamma $. Exact stationary axisymmetric rotating solutions are only known for certain values of the coupling parameter: if $\gamma =0$ one re-obtains the well known Kerr--Newman solution, and $\gamma =\sqrt{3}$ leads to the so called Kaluza--Klein rotating black hole. The latter solution, derived by a dimensional reduction of the boosted five dimensional Kerr solution to four dimensions, was first obtained in \cite{frolov} and studied in detail in \cite{horne}. Rotating solutions for arbitrary values of the coupling constant $\gamma$ were found for certain regimes: slow rotation \cite{horne} and small charge-to-mass ratio \cite{casadio}. In the context of Einstein--Maxwell theory, i.e. with $\gamma =0$, charged rotating Kaluza-Klein multi-black-hole and multi-black-string solutions were recently found \cite{matsuno}.

In this paper, we investigate the shadow cast by a Kaluza--Klein rotating dilaton black hole with charge, corresponding to a coupling parameter $\gamma =\sqrt{3}$. We adopt the observables defined in Ref. \cite{maeda} and we pay special attention to the analysis of the shadow of the Galactic supermassive black hole. In Sec. \ref{geometry}, we review the basic aspects of the geometry and we analyze the null geodesics of the rotating black hole. In Sec. \ref{bhshadow}, we obtain the size and the shape of the shadow as a function of the rotation parameter and the charge, and we study the case of Sgr A*. Finally, in Sec. \ref{conclu}, we summarize the results and discuss the observational prospects.

\section{Rotating Kaluza--Klein black hole}\label{geometry}

The rotating Kaluza--Klein solution is obtained by taking the product of the four dimensional Kerr metric, in the Boyer--Lindquist coordinates, with an extra dimension possessing translational symmetry, and then making a boost transformation with velocity $v$ along the fifth dimension. The four dimensional section of the resulting five dimensional metric has the form  \cite{frolov,horne}
\begin{equation}
ds^2  = -\frac{1-Z}{B}dt^{2}-\frac{2 a Z \sin^{2}\theta}{B\sqrt{1-v^{2}}}dt d\varphi + \left[ B(r^{2}+a^{2})+a^{2}\frac{Z}{B}\sin^{2}\theta \right] \sin^{2}\theta d\varphi^{2} + \frac{B\Sigma}{\Delta_{0}}dr^{2} + B\Sigma d\theta^{2},
\label{metric}
\end{equation}
with
\begin{equation}
B=\sqrt{1+\frac{v^{2}Z}{1-v^{2}}}, \;\;\;\;\; Z=\frac{2 m r}{\Sigma}, \;\;\;\;\; \Delta_{0} = r^2 + a^2 -2 m r , \;\;\;\;\; \Sigma=r^2+a^2 \cos^2\theta ,
\nonumber
\label{aux}
\end{equation}
where $ m $ corresponds to the mass and $a$ to the rotation parameter of the original Kerr solution. This metric, along with the non-zero components of the $U(1)$ vector field associated to $F_{\mu \nu}$
\begin{equation}
A_{t}=\frac{v}{2(1-v^{2})}\frac{Z}{B^{2}}, \;\;\;\;\; A_{\varphi}= - \frac{av}{2\sqrt{1-v^{2}}}\frac{Z}{B^{2}}\sin^{2}\theta,
\label{potencial}
\end{equation}
and the dilaton field
\begin{equation}
\phi=-\frac{\sqrt{3}}{2}\log B,
\label{dilaton}
\end{equation}
is a solution of the equations of motion of the action (\ref{emd}) for $\gamma=\sqrt{3}$. The geometry (\ref{metric}) is asymptotically flat and represents a black hole with physical mass $M$, charge $Q$, and angular momentum $J$:
\begin{equation}
M=m\left[ 1+\frac{v^{2}}{2(1-v^{2})} \right] ,
\label{M}
\end{equation}
\begin{equation}
Q=\frac{m v}{1-v^{2}},
\label{Q}
\end{equation}
and
\begin{equation}
J=\frac{m a}{\sqrt{1-v^{2}}}.
\label{J}
\end{equation}
The physical rotation parameter is defined by $A=J/M$. It is clear that the sign of the charge $Q$ is determined by the sign of the velocity $v$ of the boost, due to the physical bound $|v|<1$. Note that if $v=0$ one recovers the Kerr solution. The roots of $\Sigma$ and $\Delta_{0}$ are associated to a curvature singularity at $r=0$ and $\theta=\pi/2$, and to regular horizons, respectively. The event horizon is located at
\begin{equation}
r_{+}= m + \sqrt{m^2 - a^2},
\label{horizon}
\end{equation}
and exists if $m^{2}\geq a^{2}$; the equal sign corresponds to the extremal case. In what follows, we adopt $M=1$, which is equivalent to adimensionalize all physical quantities with the mass of the black hole. In this case, from Eq. (\ref{M}) we have that $m=2(1-v^2)/(2-v^2)$. Then, in terms of the boost velocity $v$, the presence of the event horizon requires that
\begin{equation}
|a| \leq \frac{2(1-v^2)}{2-v^2}.
\label{amax}
\end{equation}

Part of the light coming from sources behind the black hole reaches the observer after being deflected by the gravitational field of the compact object; but those photons with small impact parameters fall into the black hole. As a consequence, there is a dark zone in the sky which is named the shadow. The apparent shape of a black hole is then given by the boundary of the shadow. For obtaining this apparent shape, we need to study the geodesic structure, which is determined from the Hamilton--Jacobi equation:
\begin{equation}  \label{SHJ1}
\frac{\partial S}{\partial \lambda}=-\frac{1}{2}g^{\mu\nu}\frac{\partial S}{\partial x^{\mu}}\frac{\partial S}{\partial x^{\nu}},
\end{equation}
where $\lambda$ is an affine parameter along the geodesics, $g_{\mu\nu}$ are the components of the metric tensor, and $S$ is the Jacobi action. If the problem is \textit{separable}, the Jacobi action $S$ can be written in the form
\begin{equation}  \label{SHJ2}
S=\frac{1}{2}\mu ^2 \lambda - E t + L_z \varphi + S_{r}(r)+S_{\theta}(\theta),
\end{equation}
where $\mu $ is the mass of a test particle. The second term on the right hand side is related to the conservation of the energy $E$, while the third term is related to the conservation of the angular momentum in the direction of the axis of symmetry $L_z$. From the Hamilton--Jacobi equation, for null geodesics ($\mu  =0$) we can obtain the equations of motion for the geometry given by Eq. (\ref{metric}):
\begin{equation}
B\Sigma \frac{dt}{d\lambda}=\frac{2 m r}{\Delta_{0}}\left[(r^{2}+a^{2})\left(\frac{1}{Z}+\frac{v^{2}}{1-v^{2}} \right) E + a^{2}\sin^{2}\theta E-\frac{1}{\sqrt{1-v^{2}}}a L_z \right] ,
\label{et}
\end{equation}
\begin{equation}
B\Sigma \frac{d\varphi}{d\lambda}=\frac{2 m r}{\Delta_{0}}\left(\frac{1}{\sqrt{1-v^{2}}}a E -\frac{Z-1}{Z}\csc^{2}\theta L_z \right),
\label{ephi}
\end{equation}
\begin{equation}
B\Sigma\frac{dr}{d\lambda}=\sqrt{\mathcal{R}},
\label{er}
\end{equation}
and
\begin{equation}
B\Sigma\frac{d\theta}{d\lambda}=\sqrt{\Theta},
\label{etheta}
\end{equation}
where the functions $\mathcal{R}(r)$ and $\Theta(\theta)$ have the form
\begin{equation}
\mathcal{R}=\mathcal{R}_{\mathrm{Kerr}}+\frac{2r}{2-v^{2}}\left\{ \left[ (a E-L_z)^{2}-2 L_z^{2}-\mathcal{K}+2E^{2}r^{2}\right] v^{2}+4 a L_z E (1-\sqrt{1-v^{2}})  \right\},
\end{equation}
\begin{equation}
\Theta=\mathcal{K}+\cos^2\theta\left(a^2E^2-\frac{L_z^2}{\sin^2\theta}\right),
\end{equation}
with $\mathcal{K}$ the Carter separation constant, and
\begin{equation}
\mathcal{R}_{\mathrm{Kerr}}=\left[(r^2+a^2)E -a L_z \right]^2-(r^2-2r+a^2)\left[\mathcal{K}+(aE - L_z)^2\right].
\end{equation}
These equations determine the propagation of light in the spacetime of the Kaluza--Klein rotating dilaton black hole. As stated above, the geometry (\ref{metric}) is asymptotically flat, so the trajectories of the photons are straight lines at infinity. Light rays are characterized in general by two impact parameters, which can be expressed in terms of the constants of motion $E$, $L_z$ and $\mathcal{K}$. We define, as usual, the impact parameters for general orbits around the black hole $\xi=L_z/E$ and $\eta=\mathcal{K}/E^{2}$. We use Eq. (\ref{er}) to derive the orbits with constant $r$ in order to obtain the boundary of the shadow of the black hole. These orbits satisfy the conditions $\mathcal{R}(r)=0$ and $d\mathcal{R}(r)/dr=0$, fulfilled by the values of the impact parameters
\begin{equation}
\xi(r)=\frac{1}{a [2(1-v^{2})-r(2-v^{2})]} \left[2(a^{2}-r^{2})\sqrt{1-v^{2}} + \Delta_{0}\sqrt{r[2v^{2}+r(2-v^{2})](2-v^{2})} \right]
\label{eqxi}
\end{equation}
and
\begin{eqnarray}
\eta(r)&=&\frac{r^2}{a^2 [2(1-v^2) - r(2- v^2)]^2}  \left\{ \frac{r}{v^2-2} [8 r (5 + (r-4) r) + 4 (8 - r (31 +3(r-6)r)) v^2 \right. \nonumber \\
&&  + 2 (-32 + r (58 + 3 (r-8) r)) v^4 - (r-4)^2 (r-2) v^6] \nonumber \\
&& \left. + 4 \Delta_{0}\sqrt{r[2v^2+r(2-v^2)](2-v^2)}\sqrt{1 - v^2}
+ 2 a^2(2 v^2 - 2 v^4 + r (v^2-2)^2) \right\},
\label{eqeta}
\end{eqnarray}
which determine the contour of the shadow.

\section{Black hole shadow}\label{bhshadow}

The location of the shadow in the observer sky is better described by using the celestial coordinates $\alpha $ and $\beta$: the coordinate $\alpha$ is the apparent perpendicular distance of the image as seen from the axis of symmetry, and the coordinate $\beta$ is the apparent perpendicular distance of the image from its projection on the equatorial plane. These coordinates give the apparent position of the image in the plane that passes through the center of the black hole and is orthogonal to the line joining the observer and the black hole. We define $r_{0}$ as the observer distance to the black hole, $\theta_{0}$ as the inclination angle between the line of sight and the rotation axis of the black hole (i.e. the angular coordinate of the observer), and we can take without losing generality $\varphi _{0} =0$. Since the geometry given by Eq. (\ref{metric}) is asymptotically flat, one can place a Euclidean reference frame with the black hole at the origin, so that far away its coordinates coincide with the 
Boyer-Lindquist ones. Writing the point $(\alpha, \beta)$ in the Euclidean frame, changing to spherical coordinates, and using the geometrical description of the straight line connecting the far away observer ($r_{0} \rightarrow \infty$) with the apparent position of the image, one can find that
\begin{equation}  \label{alpha}
\alpha=\lim_{r_{0}\rightarrow \infty}\left( -r_{0}^{2}\sin\theta_{0}\frac{d\varphi}{dr}\right)
\end{equation}
and
\begin{equation}  \label{beta1}
\beta=\lim_{r_{0}\rightarrow \infty}r_{0}^{2}\frac{d\theta}{dr};
\end{equation}
for a useful diagram and a detailed calculation, see Ref. \cite{vazquez}. Using Eqs. (\ref{ephi}), (\ref{er}), and (\ref{etheta}) to calculate $d\theta /dr$ and $d\varphi /dr$, one obtains
\begin{equation}  \label{alphapsi1}
\alpha=-\xi\csc\theta_{0}
\end{equation}
and
\begin{equation}  \label{beta2}
\beta=\pm \sqrt{\eta + a^{2}\cos ^{2}\theta_{0}-\xi^{2}\cot ^{2}
\theta_{0}}.
\end{equation}
These equations, which relate the celestial coordinates with the constants of motion,  have the same form as in the case of the Kerr metric,  with the new $\xi $ and $\eta$ given by Eqs. (\ref{eqxi}) and (\ref{eqeta}) (a detailed derivation for the Kerr metric of the corresponding expressions for $\xi$ and $\eta$, and for $\alpha$ and $\beta$, can be found in \cite{vazquez}).

The size and the form of the shadow can be characterized by using the two observables introduced in \cite{maeda}. The observable $R_s$ is defined as the radius of a reference circle passing by three points of the shadow: the top position $(\alpha_t, \beta_t)$, the bottom position $(\alpha _b,\beta _b)$, and the point corresponding to the unstable retrograde circular orbit seen by an observer on the equatorial plane $(\alpha _r,0)$. The distortion parameter $\delta _{s}$ is defined by the quotient $D/R_s$, where $D$ is the difference between the endpoints of the circle and of the shadow, both of them at the opposite side of the point $(\alpha _r,0)$, i.e. corresponding to the prograde circular orbit. The radius $R_s$ gives an idea of the approximate size of the shadow, while $\delta_s$ measures its deformation with respect to the reference circle (for more details, see \cite{maeda}). If the inclination angle $\theta _{0}$ is independently known (see for example \cite{li-narayan}), precise enough measurements 
of $R_{s}$ and $\delta _{s}$ could serve to obtain the physical rotation parameter $A$ and the charge $Q$, both adimensionalized with the black hole mass as pointed out above. This information can be simply obtained by plotting the contour curves with constant $R_s$ and $\delta _s$ in the plane $(A,Q)$; the intersection point in the plane gives the corresponding values of the rotation parameter $A$ and the charge $Q$.

The gravitational effects on the shadow, which grow with $\theta _0$, are larger when the observer is situated in the equatorial plane of the black hole, that is, with an inclination angle $\theta_{0}=\pi/2$. Also, for the Galactic supermassive black hole, the inclination angle is expected to lie close to $\pi /2$. Then, we have the simple expressions
\begin{equation}  \label{alphapsi2}
\alpha=-\xi
\end{equation}
and
\begin{equation}  \label{beta3}
\beta=\pm \sqrt{\eta},
\end{equation}
for equatorial observers. The observables are then given by
\begin{equation*}
R_{s}=\frac{(\alpha _t -\alpha_r)^2 + \beta_t ^2}{2|\alpha _t -\alpha_r|},
\end{equation*}
and
\begin{equation*}
\delta _s=\frac{\tilde{\alpha}_p - \alpha_p}{R_{s}},
\end{equation*}
where $(\tilde{\alpha}_p, 0)$ and $(\alpha_p, 0)$ are the points where the reference circle and the contour of the shadow cut the horizontal axis at the opposite side of $(\alpha_r, 0)$, respectively.

\begin{figure}[t!]
\centering
\begin{minipage}{0.48\linewidth}
\centering
\includegraphics[width=0.7\linewidth]{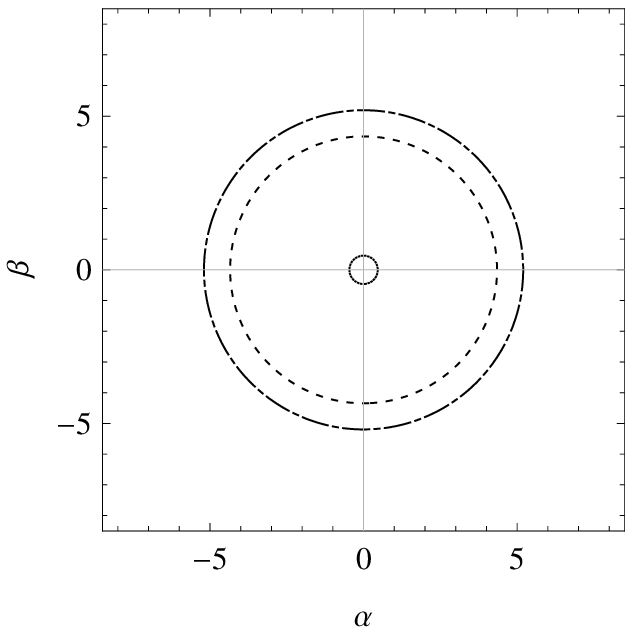}\\
\includegraphics[width=0.7\linewidth]{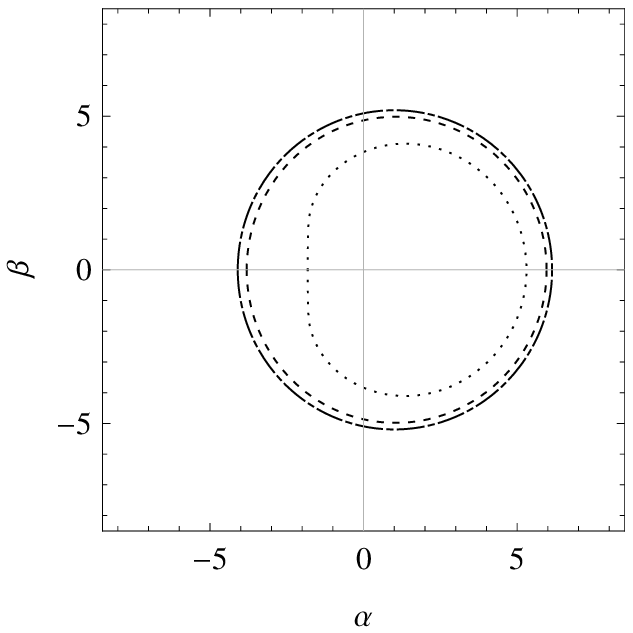}\\
\includegraphics[width=0.7\linewidth]{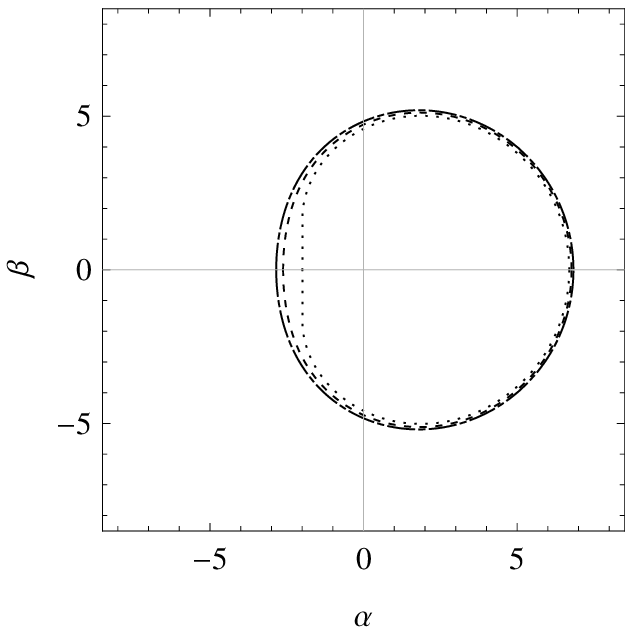}
\caption{Silhouette of the shadow cast by a black hole situated at the origin of coordinates with inclination angle $\theta _0=\pi /2$, having a physical rotation parameter $A$ and a charge $Q$. Top: $A=0$, $Q=0$ (dashed-dotted line), $0.5$ (dashed line), and $1.99$ (dotted line); center: $A=0.5$, $Q= 0$ (dashed-dotted line), $0.5$ (dashed line), and $Q_{max}=1.1298$ (dotted line); bottom:  $A=0.9$, $Q=0$ (dashed-dotted line), $0.3$ (dashed line), and $Q_{max}=0.4583$ (dotted line). All quantities were adimensionalized with the physical mass $M$ of the black hole (see text).}
\label{fig1}
\end{minipage}
\hfill
\begin{minipage}{0.48\linewidth}
\centering
\includegraphics[width=0.7\linewidth]{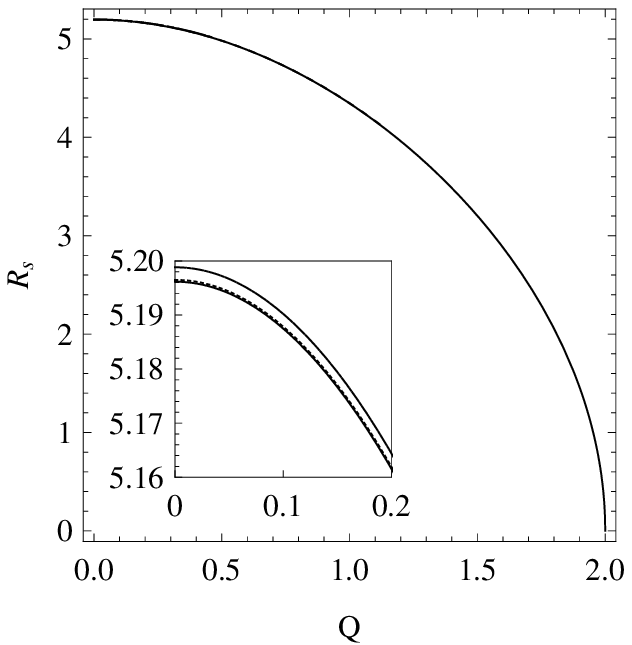}\\
\includegraphics[width=0.7\linewidth]{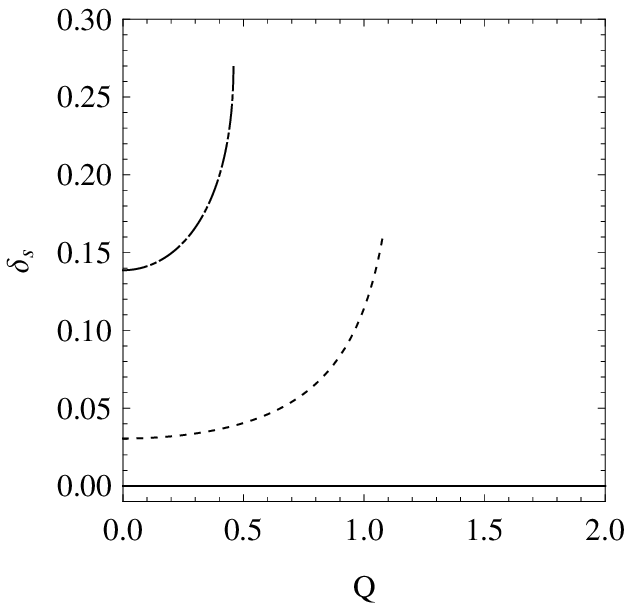}\\
\includegraphics[width=0.7\linewidth]{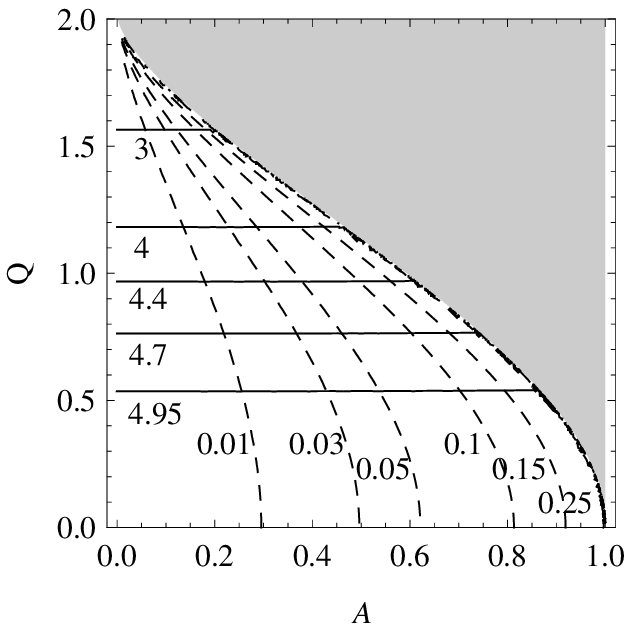}
\caption{The top and center figures show, respectively, the observables $R_{s}$ and $\delta _{s}$ as functions of the charge $Q$, for a black hole at the origin of coordinates, with an inclination angle $\theta _0=\pi /2$, and the physical spin parameters $A=0$ (full line), $A=0.5$ (dashed line), and $A=0.9$ (dashed-dotted line). A smaller range of $Q$ is shown in the frame inside the top plot, in which the different curves can be distinguished. The bottom figure shows the contour plots of $R_{s}$ (full line) and $\delta _{s}$ (dashed line) in the plane $(A,Q)$; each curve is labeled with its value of $R_{s}$ or $\delta_{s}$ (the gray zone corresponds to naked singularities).}
\label{fig2}
\end{minipage}
\end{figure}

To discuss the characteristics of the shadow, it is useful to work with the physical quantities $A$ and $Q$ (as stated above, $M=1$). As a function of these parameters, the horizon is located at
\begin{equation}
r_{+}=\frac{1}{2} \left( 3 - \sqrt{1 + 2 Q^2} + \sqrt{2}\sqrt{5 + Q^2 - 3 \sqrt{1 + 2 Q^2} - \frac{4 A^2}{1-Q^2+\sqrt{1 + 2 Q^2}}}  \right).
\label{horizAQ}
\end{equation}
Thus, in order to avoid naked singularities one needs
\begin{equation}
|A|\leq \frac{1}{2} \sqrt{2 (1 + \sqrt{1 + 2 Q^2}) - Q^2 (10 + Q^2 - 4 \sqrt{1 + 2 Q^2})},
\label{AQ}
\end{equation}
which is an equivalent condition to Eq. (\ref{amax}), and implicitly defines a maximum value for the charge $Q_{max}(A)$, as a function of $A$.

The silhouette of the black hole is defined by a contour in the $\alpha$-$\beta$ plane, which can be found implicitly using Eqs. (\ref{alphapsi2}) and (\ref{beta3}). The charge appears squared in all equations, so we take $Q\ge 0$ without losing generality. In Fig. \ref{fig1}, the shadows are shown for different values of  the physical rotation parameter $A$ and the electric charge $Q$, for a black hole located at the center of coordinates. In all the cases, we take values for the charges that go from $0$ to the critical value $Q_{max}(A)$. The non-rotating case $A=0$ is shown in the top plot, for $Q=0$ (dashed-dotted line), $Q=0.5$ (dashed line), and $Q=0.995 \, Q_{max}(0)=1.99$ (dotted line). The size of the shadow decreases with the charge, from $R_{s}=3\sqrt{3}$ until it shrinks to a point when $Q=Q_{max}(0)=2$. This is a remarkable feature of this theory when compared with the Reissner--Nordstr\"om solution of General Relativity: in the latter case the shadow fades out with the charge $Q$, starting from
$R_{s}=3\sqrt{3}$, and reaches a minimum size at $R_{s}=4$ in the extremal case $Q=1$. In the center and bottom plots we show the shadows corresponding respectively to $A=0.5$ and $A=0.9$, for $Q=0$ (center and bottom, dashed-dotted lines), $Q=0.5$ (center, dashed line), $Q=0.3$ (bottom, dashed line), $Q=Q_{max}(0.5)=1.1298$ (center, dotted line), and $Q=Q_{max}(0.9)=0.4583$ (bottom, dotted line). Again, the size of the shadows decreases with $Q$, starting from the same value of $R_{s}$ as for the Kerr solution for fixed $A$ and $Q=0$, and reaching different extremal sizes for fixed $A$ and $Q=Q_{max}(A)$, when compared with those obtained with the Kerr--Newman solution. The maximum allowed charge for fixed $A$ is larger for Kaluza--Klein black holes than for Kerr--Newman ones, so this kind of black holes can harbor larger amounts of charge than their General Relativity counterparts before becoming naked singularities. The size of the shadows of Kaluza--Klein black holes are always bigger than those of Kerr-
Newman ones, for the same values of $A$ and $Q < Q_{max}^{KN} < Q_{max}^{KK}$.

In Fig. \ref{fig2} (top plot), the behavior of the shadow radius $R_{s}$ can be seen as a function of the charge $Q$, for several values of the physical rotation parameter: $A=0$ (full line), $A=0.5$ (dotted line), and $A=0.9$ (dashed-dotted line). This observable gives information about the size of the shadow, and it is clear that it decreases with $Q$ for all $A$, and that the values of $R_{s}$ are similar for the different values of $A$ considered here, making the curves not distinguishable in the range $0<Q<2$. From the frame inside, where the range of $Q$ is smaller, it can be seen that $R_{s}$ increases with $A$, and the difference between its value for $A=0$ and for $A=0.9$, for fixed $Q$, is of order $3 \times 10^{-3}$. Each curve ends when the horizon ceases to exist and a naked singularity is formed, i.e. in the corresponding critical value $Q_{max}(A)$. The distortion of the shadow with respect to the circumference of reference $\delta _s$ is plotted in Fig. \ref{fig2} as a function of the charge
$Q$ (center plot), for $A=0$ (full line), $A=0.5$ (dotted line), and $A=0.9$ (dashed-dotted line). This observable increases with the charge until a maximum distortion, obtained for the critical value $Q_{max}(A)$. The distortion is an increasing function of $A$ for a fixed value of $Q$. For the same values of $A$ and $Q<Q_{max}^{KN}$, the shadows corresponding to Kaluza--Klein black holes are less distorted than the shadows of Kerr--Newman ones. In the bottom plot of Fig. \ref{fig2} the contour curves with constant $R_s$ and $\delta _s$ are shown in the plane $(A,Q)$, for some representative values. The gray zone represents naked singularities, which are outside the scope of the present work, and the boundary between this area and the one corresponding to black holes is given by the curve $Q_{max}(A)$. From an observational point of view this is an interesting plot, since the rotation parameter and the charge of the black hole emerge from the intersection of the curves with constant $R_s$ and $\delta _s$, 
whose values come from the observations: once those values are observed, one can extract the corresponding values of $A$ and $Q$ from this kind of plot with no ambiguity due to the fact that the contours of these observables intersect each other in a unique locus $(A,Q)$.

The observable $R_{s}$ can be used to estimate the angular size of the shadow $\theta _s = R_{s}M/D_{o}$, with $D_{o}$ the distance between the black hole and the observer.  It is not difficult to see that $\theta _s = 9.87098\times 10^{-6} R_{s}(M/M_{\odot}) ( 1 \, \text{kpc}/D_{o})$ $\mu \mathrm{as}$. To have an idea of the numbers involved, let us analyze the case of the supermassive Galactic black hole Sgr A*, for which $M=4.3 \times 10^{6}M_{\odot}$ and $D_{o}=8.3$ kpc \cite{guillessen}. The observable $\delta _s$ (\%) shows how the shadow is deformed with respect to a circle. Assuming that the observer is in the equatorial plane, the results for some values of the parameters can be summarized in the following table:
\begin{center}
\vspace{0.5cm}
\begin{tabular}{|c|c|c|c|c|c|c|}
\hline
$A=0$ & \multicolumn{3}{|c|}{$KN$}  & \multicolumn{3}{|c|}{$KKRD$}\\
\hline
$Q$ & $0$ & $0.25$ & $0.5$ & $0$ & $0.25$ & $0.5$ \\
\hline
$\theta _s (\mu \mathrm{as})$ & $26.5718$ & $26.2916$ & $25.4047$ & $26.5718$ & $26.2959$ & $25.4763$ \\
\hline
$\delta _s (\%) $  & $0$ & $0$ & $0$ & $0$ & $0$ & $0$\\
\hline \hline
$A=0.5$ & \multicolumn{3}{|c|}{$KN$}  & \multicolumn{3}{|c|}{$KKRD$}\\
\hline
$Q$ & $0$ & $0.2$ & $0.4$ & $0$ & $0.2$ & $0.4$ \\
\hline
$\theta _s (\mu \mathrm{as})$ & $26.5735$ & $26.3951$ & $25.8419$ & $26.5735$ & $26.3968$ & $25.8707$ \\
\hline
$\delta _s (\%) $  & $3.05086$ & $3.19113$ & $3.69364$ & $3.05086$ & $3.18884$ & $3.64816$\\
\hline \hline
$A=0.9$ & \multicolumn{3}{|c|}{$KN$}  & \multicolumn{3}{|c|}{$KKRD$}\\
\hline
$Q$ & $0$ & $0.05$ & $0.1$ & $0$ & $0.05$ & $0.1$ \\
\hline
$\theta _s (\mu \mathrm{as})$ & $26.5855$ & $26.5744$ & $26.5413$ & $26.5855$ & $26.5745$ & $26.5414$ \\
\hline
$\delta _s (\%) $  & $13.8666$ & $13.9301$ & $14.1248$ & $13.8666$ & $13.9300$ & $14.1236$\\
\hline
\end{tabular}
\vspace{0.5cm}
\end{center}
From the table one sees that the shadows for fixed charge and physical rotation parameter corresponding to Kaluza--Klein black holes are bigger and less distorted than those corresponding to Kerr--Newman solutions. This effect is more significant for the case $A=0.5$ and $Q=0.4$, where the difference between both values of $\theta_{s}$ is approximately 0.1\%. This means that if one hopes to see this kind of deviations from General Relativity one needs a resolution of the order of $0.01$ $\mu \mathrm{as}$ or better.

\section{Conclusions}\label{conclu}

In this article we have studied the size and the shape of the shadow cast by a rotating charged dilaton black hole, with coupling constant $\gamma =\sqrt{3}$, corresponding to a Kaluza--Klein reduction to four spacetime dimensions. We have found that, for fixed rotation parameter, mass, and charge, the presence of the dilatonic field leads to a shadow that is slightly larger and with a reduced deformation, compared with the one corresponding to the General Relativity solution.

Direct observation of black holes will be possible in the near future, and their shadows are the main feature. One way for detecting the apparent shape of a black hole is by using the very long baseline interferometer technique in (sub)millimeter wavelengths \cite{fish,Johannsen}. In particular, the Event Horizon Telescope, which consists of telescopes scattered over the Earth forming an Earth-sized high-resolution telescope \cite{EHT}, will reach a resolution of $15$ $\mu$as at $345$ GHz in the next few years \cite{lacroix}. In addition to the Event Horizon Telescope, other missions like RadioAstron and Millimetron are planned for measuring in the radio band, and MAXIM in the X-band. The first one is a space-based radio telescope launched in 2011, capable of carrying out measurements with $1$-$10$ $\mu$as angular resolution \cite{zakharov,webradio}. The space-based Millimetron mission may provide the angular resolution of $0.3$ $\mu$as or less at $0.4$ mm \cite{Johannsen}. The MAXIM project is a space-based 
X-ray interferometer with an expected angular resolution of about $0.1$ $\mu$as (see \cite{webmaxim} for further details). These instruments will be capable of observing the shadow of the supermassive Galactic black hole as well as those corresponding to nearby galaxies \cite{Johannsen}. However, in order to detect the deviations mentioned in the present article, an angular resolution of $0.01$ $\mu$as is needed. So it will not be until the next generation of instruments that it will be possible to distinguish between the different gravitational theories considered here.

\begin{acknowledgments}

This work was supported by CONICET and Universidad de Buenos Aires.

\end{acknowledgments}


\begin{thebibliography}{99}

\bibitem{darwin} C. Darwin, Proc. Roy. Soc London A \textbf{249}, 180 (1959).

\bibitem{otros} J.-P. Luminet, Astron. Astrophys. \textbf{75}, 228 (1979);
H.C. Ohanian, Am. J. Phys. \textbf{55}, 428 (1987); R.J. Nemiroff, Am. J.
Phys. \textbf{61}, 619 (1993); V. Bozza, S. Capozziello, G. Iovane, and G.
Scarpetta, Gen. Relativ. Gravit. \textbf{33}, 1535 (2001).

\bibitem{eiroto} E.F. Eiroa, G.E. Romero, and D.F. Torres, Phys. Rev. D
\textbf{66}, 024010 (2002);  E.F. Eiroa and D.F. Torres, Phys. Rev. D
\textbf{69}, 063004 (2004).

\bibitem{boz} V. Bozza, Phys. Rev. D \textbf{66}, 103001 (2002).

\bibitem{numerical} K.S. Virbhadra and G.F.R. Ellis, Phys. Rev. D \textbf{62}, 084003 (2000); K.S. Virbhadra and C.R. Keeton, Phys. Rev. D \textbf{77}, 124014 (2008); K.S. Virbhadra, Phys. Rev. D \textbf{79}, 083004 (2009).

\bibitem{alternative}  K.S. Virbhadra, D. Narasimha, and S.M. Chitre, Astron. Astrophys. \textbf{337}, 1 (1998); A. Bhadra, Phys. Rev. D  \textbf{67}, 103009 (2003); E.F. Eiroa, Phys. Rev. D \textbf{73}, 043002 (2006); K. Sarkar and A. Bhadra, Class. Quantum Grav. \textbf{23}, 6101 (2006); N. Mukherjee and A.S. Majumdar, Gen. Relativ. Gravit. \textbf{39}, 583 (2007); G. N. Gyulchev and S. S. Yazadjiev, Phys. Rev. D \textbf{75}, 023006 (2007); S. Chen and J. Jing, Phys. Rev. D \textbf{80}, 024036 (2009); Y. Liu, S. Chen and J. Jing, Phys. Rev. D \textbf{81}, 124017  (2010); E.F. Eiroa and C.M. Sendra, Class. Quantum Grav. \textbf{28}, 085008 (2011).

\bibitem{bwlens} E.F. Eiroa, Phys. Rev. D \textbf{71}, 083010 (2005); R. Whisker, Phys. Rev. D \textbf{71}, 064004 (2005); A.S. Majumdar and N. Mukherjee, Int. J. Mod. Phys. D \textbf{14}, 1095 (2005); C. R. Keeton and A. O. Petters, Phys. Rev. D \textbf{73}, 104032 (2006); E.F. Eiroa, Phys. Rev. D \textbf{73}, 043002 (2006);  S. Pal and S. Kar, Class. Quantum Grav. \textbf{25}, 045003 (2008); A.Y. Bin-Nun, Phys. Rev. D \textbf{81}, 123011 (2010);  A.Y. Bin-Nun, Phys. Rev. D \textbf{82}, 064009 (2010); E.F. Eiroa and C.M. Sendra, Phys. Rev. D  \textbf{86}, 083009 (2012).

\bibitem{bozza1} V.Bozza, Phys. Rev. D \textbf{67}, 103006 (2003); V. Bozza, F. De Luca, G. Scarpetta, and M. Sereno, Phys. Rev. D \textbf{72}, 083003 (2005); V. Bozza, F. De Luca, and G. Scarpetta, Phys. Rev. D \textbf{74}, 063001 (2006).

\bibitem{vazquez} S. V\'azquez and E. Esteban, Nuovo Cim. \textbf{119B}, 489 (2004).

\bibitem{bozza2} V. Bozza and G. Scarpetta, Phys. Rev. D \textbf{76}, 083008 (2007).

\bibitem{kraniotis} G.V. Kraniotis, Class. Quantum Grav. \textbf{28}, 085021 (2011).

\bibitem{bardeen} J. Bardeen, \textit{Black Holes}, \'Ecole d' \'et\'e de Physique Th\'eorique, Les Houches 1972, edited by C. De Witt and B.S. De Witt (Gordon and Breach Science Publishers, New York, 1973).

\bibitem{chandra} S. Chandrasekhar, \textit{The Mathematical Theory of Black Holes} (Oxford University Press, New York, 1992).

\bibitem{falcke} H. Falcke, F. Melia, and E. Agol, Astrophys. J. \textbf{528}, L13 (2000).

\bibitem{devries} A. de Vries, Class. Quant. Grav. \textbf{17}, 123 (2000).

\bibitem{takahashi} R. Takahashi, Astrophys. J. \textbf{611}, 996 (2004).

\bibitem{hioki} K. Hioki and U. Miyamoto, Phys. Rev. D \textbf{78}, 044007
(2008).

\bibitem{bambi} C. Bambi and K. Freese, Phys. Rev. D \textbf{79}, 043002 (2009).

\bibitem{maeda} K. Hioki and K.I. Maeda, Phys. Rev. D \textbf{80}, 024042
(2009).

\bibitem{schee1} J. Schee and Z. Stuchlik, Int. Jour. Mod. Phys. D \textbf{18}, 983 (2009).

\bibitem{schee2} Z. Stuchlik and J. Schee, Class. Quantum Grav. \textbf{27},  215017 (2010).

\bibitem{amarilla1} L. Amarilla, E.F. Eiroa, and G. Giribet, Phys. Rev. D \textbf{81}, 124045 (2010).

\bibitem{amarilla2} L. Amarilla and E.F. Eiroa, Phys. Rev. D \textbf{85}, 064019 (2012).

\bibitem{yumoto} A. Yumoto, D. Nitta, T. Chiba, and N. Sugiyama, Phys. Rev. D \textbf{86}, 103001 (2012).

\bibitem{zakharov} A.F. Zakharov, A.A. Nucita, F. De Paolis, and G. Ingrosso, New Astron. \textbf{10}, 479 (2005); A.F. Zakharov, F. De Paolis, G. Ingrosso, and A.A. Nucita, Astron. Astrophys. \textbf{442}, 795 (2005); F. De Paolis, G. Ingrosso, A.A. Nucita, A. Qadir, and A.F. Zakharov, Gen. Relativ. Gravit. \textbf{43}, 977 (2011).

\bibitem{morris} M. R. Morris, L. Meyer, and A. M. Ghez, Res. Astron.
Astrophys. \textbf{12}, 995 (2012).

\bibitem{lacroix} T. Lacroix and J. Silk, arXiv:1211.4861 (2012).

\bibitem{bozzareview} V. Bozza, Gen. Relativ. Gravit. \textbf{42}, 2269 (2010).

\bibitem{mae} G.W. Gibbons and K. Maeda, Nucl. Phys. \textbf{B298}, 741 (1988).

\bibitem{ghs} D. Garfinkle, G.T. Horowitz, and A. Strominger, Phys. Rev. D
\textbf{43}, 3140 (1991); \textit{ibid.} \textbf{45}, 3888(E) (1992).

\bibitem{horne} J.H. Horne and G.T. Horowitz, Phys. Rev. D \textbf{46}, 1340 (1992).

\bibitem{frolov} V. Frolov, A. Zelnikov, and U. Bleyer, Ann. Phys. (Leipzig) \textbf{499}, 371 (1987).

\bibitem{casadio} R. Casadio, B. Harms, Y. Leblanc, and P.H. Cox, Phys. Rev. D \textbf{55}, 814 (1997).

\bibitem{matsuno} K. Matsuno, H. Ishihara, M. Kimura, and T. Tatsuoka, Phys. Rev. D \textbf{86}, 104054 (2012).

\bibitem{li-narayan} L.-X. Li, R. Narayan, and J.E. McClintock, Astrophys. J. \textbf{691}, 847 (2009).

\bibitem{guillessen} S. Guillessen, F. Eisenhauer, S. Trippe, T. Alexander, R. Genzel, F. Martins, and T. Ott, Astrophys. J. \textbf{692}, 1075 (2009).

\bibitem{fish} V.L. Fish and S.S. Doeleman, in Proceedings IAU Symposium No. 261 (2009).

\bibitem{Johannsen} T. Johannsen, D. Psaltis, S. Gillessen, D.P. Marrone, F. \"Ozel, S.S. Doeleman, and V.L. Fish, Astrophys. J. \textbf{758}, 30 (2012).

\bibitem{EHT}  http://www.eventhorizontelescope.org.

\bibitem{webradio} http://www.asc.rssi.ru/radioastron.

\bibitem{webmaxim} http://maxim.gsfc.nasa.gov.

\end{thebibliography}
\end{document}